\documentclass[twocolumn,showpacs,preprintnumbers,amsmath,amssymb,pra,aps,superscriptaddress]{revtex4-1}
\usepackage[space]{grffile}
\usepackage[]{hyperref}
\usepackage[capitalize]{cleveref}
\usepackage{tikz}
\usetikzlibrary{calc, positioning}

\newcommand{\ket}[1]{\left\lvert #1 \right\rangle}
\newcommand{\bra}[1]{\left\langle #1 \right\rvert}
\newcommand{\Tone}{T_{1}}

\newcommand{\Hz}{\mathrm{Hz}}
\newcommand{\kHz}{\mathrm{kHz}}

\newcommand{\s}{\mathrm{s}}
\newcommand{\us}{\mu\mathrm{s}}
\newcommand{\ns}{\mathrm{ns}}
\newcommand{\ms}{\mathrm{ms}}

\newcommand{\dB}{\mathrm{dB}}

\newcommand{\errfrac}{\varepsilon}
\newcommand{\er}{\varepsilon_\mathrm{R}}
\newcommand{\ec}{\varepsilon_\mathrm{C}}

\newcommand{\FCl}{F_\mathrm{Cl}}
\newcommand{\FCla}{\FCl^\mathrm{a}}
\newcommand{\FClb}{\FCl^{\mathrm{b}}}
\newcommand{\FClGST}{F_\mathrm{Cl}^\mathrm{GST}}

\newcommand{\FClTone}{F_\mathrm{Cl}^{(\Tone)}}

\newcommand{\tauconv}{\tau}
\newcommand{\taup}{\tau_{\mathrm{p}}}
\newcommand{\tauc}{\tau_{\mathrm{c}}}

\newcommand{\NCl}{N_{\mathrm{Cl}}}
\newcommand{\Clgroup}{\mathcal{G}_\mathrm{Cl}}

\newcommand{\AG}{A_\mathrm{G}}
\newcommand{\AD}{A_\mathrm{D}}

\newcommand{\SD}{S_{\Tone}}

\newcommand{\ps}{p_\mathrm{s}}
\newcommand{\pc}{p_\mathrm{Cl}}

\newcommand{\var}{\mathrm{var}}
\newcommand{\covar}{\mathrm{covar}}
\newcommand{\NIT}{N_{\mathrm{it}}}

\newcommand{\gates}{\{G\}}

\newcommand{\pe}{p_\mathrm{e}}
\newcommand{\pcsupp}{p_{\mathrm{c}}}
\newcommand{\pej}{p_\mathrm{e}^{(j)}}
\newcommand{\psj}{p_\mathrm{s}^{(j)}}
\newcommand{\ppulse}{p_{\mathrm{pulse}}}
\newcommand{\pctone}{p_{\mathrm{c}}^{(\Tone)}}
\newcommand{\psc}{p_{\mathrm{s}}^{(\mathrm{c})}}
\newcommand{\pszero}{p_{\mathrm{s}}^{(0)}}
\newcommand{\psone}{p_{\mathrm{s}}^{(1)}}

\newcommand{\tro}{\tau_\mathrm{RO}}
\newcommand{\tcl}{\tau_\mathrm{Cl}}
\newcommand{\trrb}{\tau_\mathrm{RRB}}
\newcommand{\tm}{\tau_\mathrm{m}}
\newcommand{\ta}{\tau_\mathrm{a}}
\newcommand{\tb}{\tau_\mathrm{b}}

\begin{document}
\title{Restless Tuneup of High-Fidelity Qubit Gates}

\author{M.~A.~Rol}
\author{C.~C.~Bultink}
\affiliation{QuTech, Delft University of Technology, P.O. Box 5046, 2600 GA Delft, The Netherlands}
\affiliation{Kavli Institute of Nanoscience, Delft University of Technology, P.O. Box 5046, 2600 GA Delft, The Netherlands}
\author{T.~E.~O'Brien}
\affiliation{Instituut-Lorentz for Theoretical Physics, Leiden University, Leiden, The Netherlands}
\author{S.~R.~de~Jong}
\affiliation{QuTech, Delft University of Technology, P.O. Box 5046, 2600 GA Delft, The Netherlands}
\affiliation{Kavli Institute of Nanoscience, Delft University of Technology, P.O. Box 5046, 2600 GA Delft, The Netherlands}
\author{L.~S.~Theis}
\affiliation{Theoretical Physics, Saarland University, 66123 Saarbr¨ucken, Germany}
\author{X.~Fu}
\affiliation{QuTech, Delft University of Technology, P.O. Box 5046, 2600 GA Delft, The Netherlands}
\author{F.~Luthi}
\affiliation{QuTech, Delft University of Technology, P.O. Box 5046, 2600 GA Delft, The Netherlands}
\affiliation{Kavli Institute of Nanoscience, Delft University of Technology, P.O. Box 5046, 2600 GA Delft, The Netherlands}
\author{R.~F.~L.~Vermeulen}
\affiliation{QuTech, Delft University of Technology, P.O. Box 5046, 2600 GA Delft, The Netherlands}
\affiliation{Kavli Institute of Nanoscience, Delft University of Technology, P.O. Box 5046, 2600 GA Delft, The Netherlands}
\author{J.~C.~de~Sterke}
\affiliation{Topic Embedded Systems B.V., P.O. Box 440, 5680 AK Best, The Netherlands}
\affiliation{QuTech, Delft University of Technology, P.O. Box 5046, 2600 GA Delft, The Netherlands}
\author{A.~Bruno}
\affiliation{QuTech, Delft University of Technology, P.O. Box 5046, 2600 GA Delft, The Netherlands}
\affiliation{Kavli Institute of Nanoscience, Delft University of Technology, P.O. Box 5046, 2600 GA Delft, The Netherlands}
\author{D.~Deurloo}
\affiliation{Netherlands Organisation for Applied Scientific Research (TNO),  P.O. Box 155, 2600 AD Delft, The Netherlands}
\affiliation{QuTech, Delft University of Technology, P.O. Box 5046, 2600 GA Delft, The Netherlands}
\author{R.~N.~Schouten}
\affiliation{QuTech, Delft University of Technology, P.O. Box 5046, 2600 GA Delft, The Netherlands}
\affiliation{Kavli Institute of Nanoscience, Delft University of Technology, P.O. Box 5046, 2600 GA Delft, The Netherlands}
\author{F.~K.~Wilhelm}
\affiliation{Theoretical Physics, Saarland University, 66123 Saarbr¨ucken, Germany}
\author{L.~DiCarlo}
\affiliation{QuTech, Delft University of Technology, P.O. Box 5046, 2600 GA Delft, The Netherlands}
\affiliation{Kavli Institute of Nanoscience, Delft University of Technology, P.O. Box 5046, 2600 GA Delft, The Netherlands}

\date{\today}

\begin{abstract}
We present a tuneup protocol for qubit gates with tenfold speedup over traditional methods reliant on qubit initialization by energy relaxation.
This speedup is achieved by constructing a cost function for Nelder-Mead optimization from real-time correlation of non-demolition measurements interleaving gate operations without pause.
Applying the protocol on a transmon qubit achieves 0.999 average Clifford fidelity in one minute, as independently verified using randomized benchmarking and gate set tomography.
The adjustable sensitivity of the cost function allows detecting fractional changes in gate error with nearly constant signal-to-noise ratio.
The restless concept demonstrated can be readily extended to the tuneup of two-qubit gates and measurement operations.
\end{abstract}

\maketitle
Reliable quantum computing requires the building blocks of algorithms, quantum gates, to be executed with low error. Strategies aiming at quantum supremacy without error correction~\cite{Boixo16,Dallaire-Demers16} require $\sim10^3$ gates, and thus gate errors $\sim10^{-3}$. Concurrently, a convincing demonstration of quantum fault tolerance using the circuits Surface-17 and -49~\cite{Horsman12, Tomita14} under development by several groups worldwide requires gate errors one order of magnitude below the $\sim10^{-2}$ threshold of surface code~\cite{Fowler12, Martinis15}.

The quality of qubit gates depends on qubit coherence times and the accuracy and precision of the pulses realizing them. With the exception of a few systems known with metrological precision~\cite{Anderson15}, pulsing requires meticulous calibration by closed-loop tuning, i.e., pulse adjustment based on experimental observations. Numerical optimization algorithms have been implemented to solve a wide range of tuning problems with a cost-effective number of iterations~\cite{Egger14,Kelly14, Kelly16, McClure16, Bultink16, Cerfontaine16}.
However, relatively little attention has been given to quantitatively exploring the speed and robustness of the algorithms used.
This becomes crucial with more complex and precise quantum operations, as the number of parameters and requisite precision of calibration grow.

Though many aspects of tuning qubit gates are implementation independent, some details are specific to physical realizations.
Superconducting transmon qubits are a promising hardware for quantum computing, with gate times already exceeding coherence times by three orders of magnitude. Conventional gate tuneup relies on qubit initialization, performed passively by waiting several times the qubit energy-relaxation time $\Tone$ or actively through feedback-based reset~\cite{Riste12b}. Passive initialization becomes increasingly inefficient as $\Tone$ steadily increases~\cite{Devoret13, Wang15}, while feedback-based reset is technically involved~\cite{Riste15b}.

In this Letter, we present a gate tuneup method that dispenses with $\Tone$ initialization and achieves tenfold speedup over the state of the art~\cite{Kelly14} without active reset. Restless tuneup exploits the real-time correlation of quantum-non-demolition (QND) measurements interleaving gate operations without pause, and the evaluation of a cost function for numerical optimization with adjustable sensitivity at all levels of gate fidelity.  This cost function is obtained from a simple modification of the gate sequences of conventional randomized benchmarking (CRB) to penalize both gate errors within the qubit subspace and leakage from it. We quantitatively match the signal to noise ratio of this cost function with a model that includes measured $\Tone$ fluctuations.
Restless tuneup robustly achieves $\Tone$-dominated gate fidelity of $0.999$,
verified using both CRB with $\Tone$ initialization and a first implementation of gate set tomography (GST) in a superconducting qubit.
While this performance matches that of conventional tuneup, restless is tenfold faster and converges in one minute.

\begin{figure}
  \centering
    \includegraphics{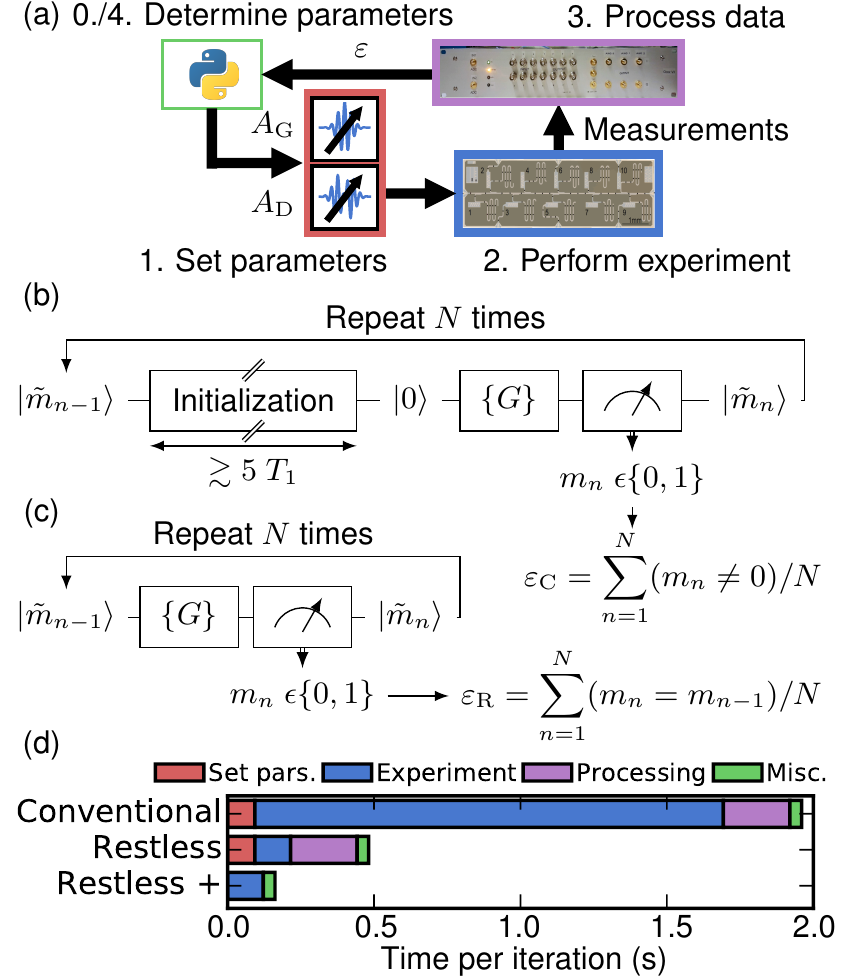}
  \caption{\label{fig:concept}
  (a) A general qubit gate tuneup loop.
  In conventional tuneup (b), the qubit is initialized before measuring the effect of $\gates$.
  In restless tuneup (c), the qubit is not initialized but $m_{n-1}$ is used to estimate the initial state ($\ket{\tilde{m}_{n-1}}$).
  (d)
  Benchmark of various contributions to the time per iteration in conventional and restless tuneup, without and with technical improvements (see text for details).
  }
\end{figure}

In many tuneup routines~[\cref{fig:concept}(a)], the relevant information from the measurements can be expressed as the fraction $\errfrac$ of non-ideal outcomes ($m_n$).
In conventional gate tuneup, a qubit is repeatedly initialized in the ground state $\ket{0}$, driven by a set of gates ($\gates$) whose net operation is ideally identity, and measured [\cref{fig:concept}(b)].
The conventional cost function is the raw infidelity,
\begin{equation*}
\ec = \sum_{n=1}^N (m_n \neq 0)/N.
\end{equation*}

The central idea of restless tuning [\cref{fig:concept}(c)] is to remove the time-costly initialization step by measuring the correlation between subsequent QND measurements interleaving gate operations without any rest~\footnote{except $3.25~\us$ needed for passive depletion of photons leftover from the $1~\us$ measurement~\cite{Bultink16}}.
For example, when the net ideal gate operation is a bit flip, we can define the error fraction
\begin{equation}
\er = \sum_{n=2}^N (m_n = m_{n-1})/N.
\end{equation}

We demonstrate restless tuneup of DRAG pulses~\cite{Motzoi09} on the transmon qubit recently reported in~\cite{Bultink16}. We choose DRAG pulses (duration $\taup=20~\ns$) for their proven ability to reduce gate error and leakage~\cite{Chow10b,Chen16b} with few-parameter analytic pulse shapes, consisting of Gaussian (G) and derivative of Gaussian (D) envelopes of the in- and quadrature-phase components of a microwave drive at the transition frequency $f$ between qubit levels $\ket{0}$ and $\ket{1}$.
These components are generated using four channels of an arbitrary waveform generator (AWG), frequency upconversion by sideband modulation of one microwave source, and two I-Q mixers.
The G and D components are combined inside a vector switch matrix (VSM)~\cite{Asaad16} (details in~\cite{SOMrestless}).
A key advantage of this scheme using four channels is the ability to independently set the G and D amplitudes ($\AG$ and $\AD$, respectively), without uploading new waveforms to the AWG.

To measure the speedup obtained from the restless method, we must take the complete iteration into account.
The traditional iteration of a tuneup routine involves:
(1) setting parameters (4 channel amplitudes on a Tektronix 5014 AWG);
(2) acquiring $N=8000$ measurement outcomes;
(3) sending the measurement outcomes to the computer and processing them;
and (4) miscellaneous overhead that includes determining the parameters for the next iteration, as well as saving and plotting data.
In \cref{fig:concept}(d), we visualize these costs for an example optimization experiment.
We intentionally penalize the restless method by choosing a large number of gates ($\sim550$).
Even in these conditions, restless sequences reduce the acquisition time from $1.60$ to $0.12~\s$.
However, the improvement in total time per iteration (from $1.98$ to $0.50~\s$) is modest due to $0.38~\s$ of overhead.

We take two steps to reduce overhead. The $0.23~\s$ required to send all measurement outcomes to the computer and then calculate the error fraction is reduced to $<1~\ms$ by calculating the fraction in real time using the same FPGA system that digitizes and  processes the raw measurement signals into bit outcomes. The $0.09~\s$ required to set the four channel amplitudes in the AWG is reduced to $1~\ms$ by setting $\AG$ and $\AD$ in the VSM.
With these two technical improvements, the remaining overhead is dominated by the miscellaneous contributions ($40~\ms$).
This reduces the total time per restless (conventional) iteration to $0.16~\s$ ($1.64~\s$).

A quantity of common interest in gate tuneup is the average Clifford fidelity $\FCl$, which is typically measured using CRB.
In CRB, $\gates$ consists of sequences of $\NCl$ random Cliffords, including a final recovery Clifford that makes the ideal net operation identity.
Following~\cite{Epstein14}, we compose the 24 Cliffords from the set of $\pi$ and $\pm \pi /2$ rotations around the $x$ and $y$ axes, which requires an average of 1.875 gates per Clifford. Gate errors make $\ec$ increase with $\NCl$ as~\cite{Magesan11, Magesan12}
\begin{equation}
\label{eq:CRB}
1-\ec = A \cdot  (\pc) ^{\NCl} + B.
\end{equation}
Here, $A$ and $B$ are constants determined by state preparation and measurement error (SPAM), and $1-\pc$ is the average depolarizing probability per gate, making $\FCl =\frac{1}{2}+\frac{1}{2} \pc$.
Extracting $\FCl$ from a CRB experiment involves measuring $\ec$ for different $\NCl$ and fitting \cref{eq:CRB}.
However, for tuning it is sufficient to optimize $\ec$  at one choice of $\NCl$, because $\ec(\NCl)$ decreases monotonically with $\FCl$~\cite{Kelly14}.

Due to leakage,  CRB sequences and $\ec$ are not well suited for restless tuneup.
Typically, there is significant overlap in readout signals for the first- ($\ket{1}$) and second- ($\ket{2}$) excited state of a transmon.
A transmon in $\ket{2}$ can produce a string of identical measurement outcomes until it relaxes back to the qubit subspace.
If the ideal net operation of $\gates$ is identity, the measurement outcomes can be indistinguishable from ideal behavior.
By choosing the recovery Clifford for restless randomized benchmarking (RRB) sequences so that the ideal net operation of $\gates$ is a bit flip, we penalize leakage and make $\er$ a suitable cost function.

\begin{figure}
  \centering
  \includegraphics{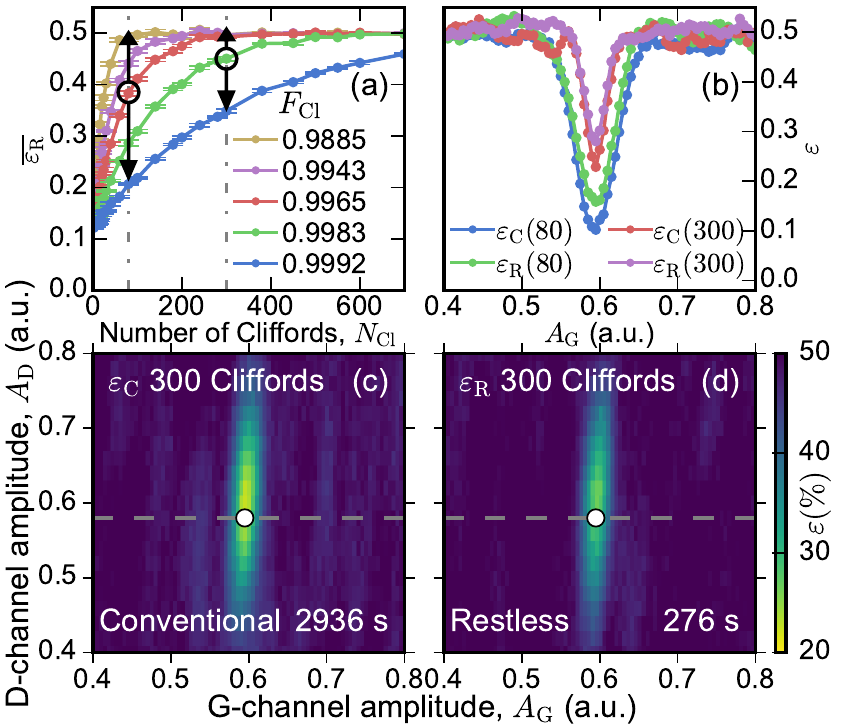}
  \caption{
  (a) Average error fraction of RRB for different $\FCl$ vs $\NCl$.
  (b) $\ec$ and $\er$ as a function of $\AG$ for $\NCl=80$ and $\NCl=300$.
  The curves are denoted by a dashed line in (c-d).
  (c-d) $\errfrac$ for $\NCl = 300$ as a function of $\AG$ and $\AD$.
  White circles indicate minimal $\errfrac$.
  Total acquisition time is shown at the bottom right.
  }
\label{fig:equivalence_cost_func}
\end{figure}

We now examine the suitability of the restless scheme for optimization (\cref{fig:equivalence_cost_func}).
Plots of the average $\er(\NCl)$ [$\overline{\er}(\NCl)$] at various $\FCl$ (controlled via $\AG$) behave similarly to $\ec$ in CRB.
Furthermore, $\er$ is minimized at the same $\AG$ as $\ec$, with only a shallower dip because of SPAM. The ($\AG$, $\AD$) landscapes for both cost functions [\cref{fig:equivalence_cost_func}(c-d)] are smooth around the optimum, making them suitable for numerical optimization. The fringes far from the optimum arise from the limited number of seeds (always $200$) used to generate the RB sequences.
Note that, while the landscapes are visually similar, the difference in time required to map them is striking,  $\sim50~\min$ for $\ec$ versus $<5~\min$ for $\er$ at $\NCl=300$.

\begin{figure}
  \centering
  \includegraphics{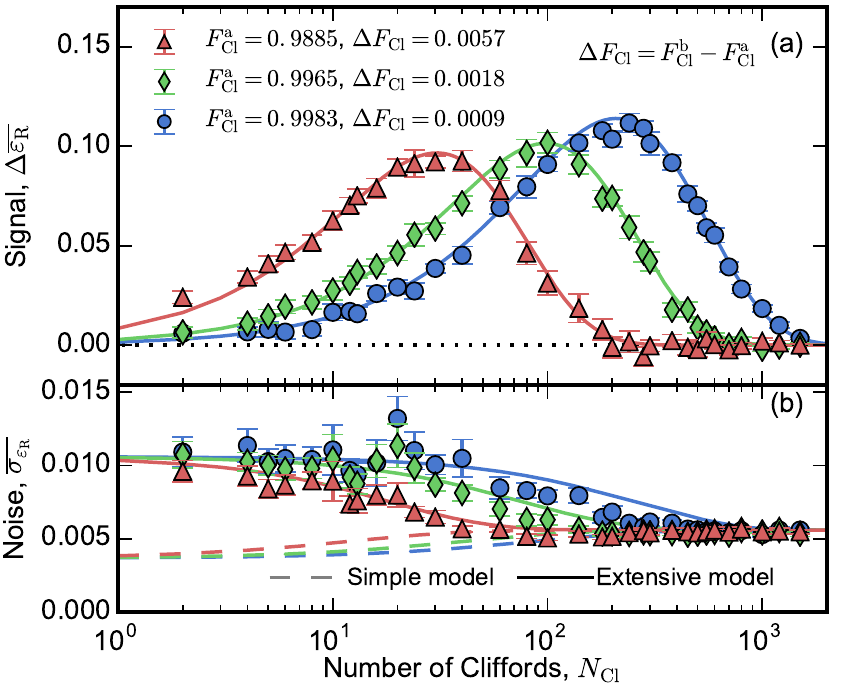}
  \caption{
  (a) Signal $\Delta \overline{\er}$ for a halving of the gate infidelity, plotted as a function $\NCl$ at $\FCla \sim 0.989$ (red), $0.996$ (green) and $0.998$ (blue).
  (b) Noise dependence on $\NCl$ at the same fidelity levels.
  Added curves are obtained from the two models described in the main text.
  }
\label{fig:signal_noise}
\end{figure}

The sensitivity of $\er$ to the tuning parameters depends on both the gate fidelity and $\NCl$.
This can be seen in the variations between curves in \cref{fig:equivalence_cost_func}(a). In order to quantify this sensitivity, we define a signal-to-noise ratio (SNR).
For signal we take the average change in the error fraction, $\Delta \overline{\er} = \overline{\er}(\FClb) - \overline{\er}(\FCla)$, from $\FCla$ to $\FClb\approx{\frac{1}{2}+\frac{1}{2}\FCla}$ (halving the infidelity).
For noise we take $\overline{\sigma_{\er}}$, the average standard deviation of $\er$ between $\FCla$ and $\FClb$.
We find that the maximal SNR remains $\sim15$ for an optimal choice of $\NCl$ that increases with $\FCla$ (\cref{fig:signal_noise} and details in~\cite{SOMrestless}).
This allows tuning in logarithmic time since reducing error rates $p\rightarrow p/2^M$ requires only $M$ optimization steps.

A simple model describes the measurement outcomes as independent and binomially distributed with error probability $\er$, as per \cref{eq:CRB} with $\ec\rightarrow\er$.
This model captures all the essential features of the signal. However, it only quantitatively matches the noise at high $\NCl$.
Experiment shows an increase in noise at low $\NCl$. In this range, $\er$ is dominated by SPAM, which is primarily due to $\Tone$.
We surmise that the increase stems from $\Tone$ fluctuations~\cite{Muller15} during the acquisition of statistics in these RRB experiments.
To test this hypothesis, we develop an extensive model incorporating $\Tone$ fluctuations into the calculation of both signal and noise~\cite{SOMrestless}.
We find good agreement with experimental results using independently measured values of $\overline{\Tone}$ and $\sigma_{\Tone}$.

\begin{figure}[t!]
  \centering
  \includegraphics{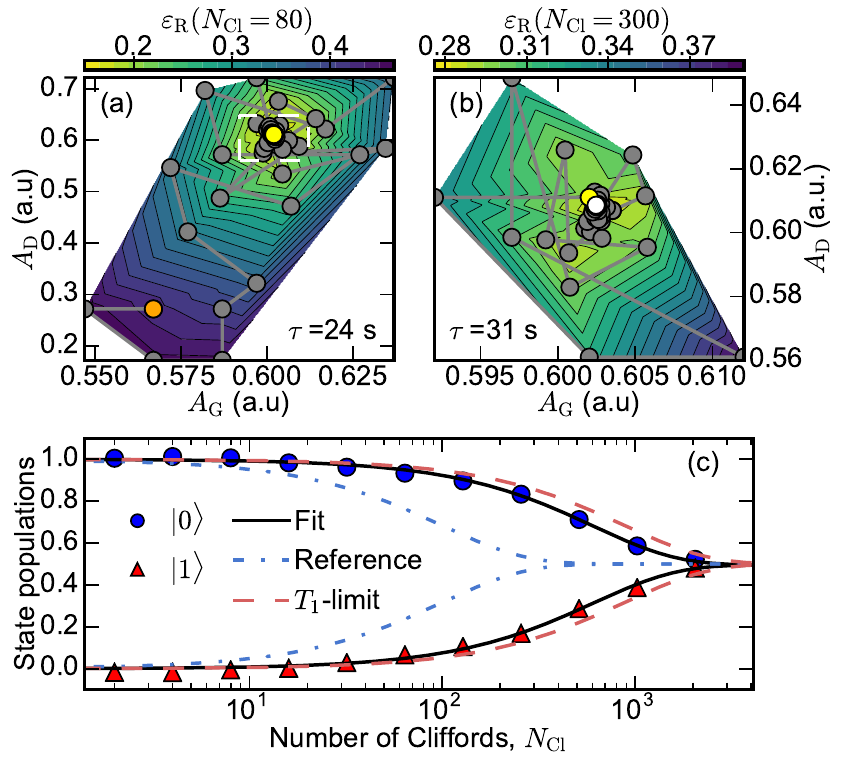}
  \caption{
  Two-parameter restless tuneup using a two-step optimization, first at $\NCl = 80$ (a) and then at $\NCl = 300$ (b).
  Contour plots show a linear interpolation of $\er$.
  The starting point, intermediate result and final result are marked by orange, yellow and white dots, respectively.
  (c) CRB of tuned pulses ($\FCl=0.9991$), and compared to $\FClTone = 0.9994$ and  $\FCl=0.995$ for reference.
  }
\label{fig:optimization}
\end{figure}

Following its validation, we now employ $\er$ in a two-step numerical optimization protocol (\cref{fig:optimization}).
We choose the Nelder-Mead algorithm~\cite{Nelder65} as it is derivative-free and easy to use, requiring only the specification of a starting point and initial stepsizes.
The first step using $\er(\NCl=80)$ ensures convergence even when starting relatively far from the optimum, while the second step using $\er(\NCl=300)$ fine tunes the result. We test the optimization for four realistic starting deviations from the optimal parameters $(\AD^{\text{opt}},\AG^{\text{opt}})$.
$\AG$ starts at roughly $6\%$ above or below $\AG^{\text{opt}}$, chosen as a worst-case estimate from a Rabi-oscillation experiment. $\AD$ starts at roughly half or double $\AD^{\text{opt}}$.
The initial stepsizes are $\Delta\AG\approx-0.03\AG^{\text{opt}}$, $\Delta\AD\approx-0.25\AD^{\text{opt}}$ for the first step, and $\Delta\AG\approx-0.01\AG^{\text{opt}}$, $\Delta\AD\approx-0.08\AD^{\text{opt}}$ for the second step.

We assess the accuracy of the above optimization and compare to traditional methods.
A CRB experiment [\cref{fig:optimization}(c)] following two-parameter restless optimization indicates $\FCl=0.9991$.
This value matches the average achieved by both restless and conventional tuneups for the different starting conditions.
We also implement GST to independently verify results obtained using CRB.
From the process matrices we extract the average GST Clifford fidelity, $\FClGST=0.99907\pm 0.00003$ ($0.99909\pm 0.00003$) for restless (conventional) tuneup~\cite{SOMrestless}, consistent with the value obtained from CRB.

\begin{table}[!t]
    \begin{tabular}{| c| c | c | c | c |}
    \hline
    \hline
    &\multicolumn{2}{c |}{{2-par. ($\AG$, $\AD$)}} & \multicolumn{2}{c|}{{3-par. ($\AG$, $\AD$, $f$)}}\\ \cline{2-5}
    & {conv.} & {restl.} & {conv.} & {restl.} \\ \hline
    $\overline{\FCl}$ &0.9991 & 0.9991 & 0.9990 & 0.9990\\ \hline
    $\sigma_{\FCl}$ & $3\cdot10^{-5}$ &  $3\cdot10^{-5}$ & 0.0001 & 0.0001\\ \hline
    $\overline{\tauconv}$ & $660~\s$ & $59~\s$ & $610~\s$ & $66 ~\s$ \\ \hline
    $\sigma_{\tauconv}$ & $110~\s$ & $11~\s$ & $110~\s$ & $13~\s$ \\ \hline
    $\overline{\NIT}$ & $400$ & $370$ & $370$ & $420$ \\ \hline
    $\sigma_{\NIT}$ & $70$ & $70$ & $70$ & $80$ \\ \hline
    $\overline{\FClTone}$ &\multicolumn{2}{c |}{0.9994} &\multicolumn{2}{c|}{0.9993} \\ \hline
    $\overline{\Tone}$ & \multicolumn{2}{c |}{$21.4~\us$} & \multicolumn{2}{c|}{$19.3~\us$}\\ \hline \hline
    \end{tabular}
    \caption{
    \label{tab:verification}
    Tuning protocol performance. Mean (overlined) and standard deviations (denoted by $\sigma$) of $\FCl$, time to convergence $\tauconv$, and number of iterations $\NIT$ for restless and conventional tuneups with 2 and 3 parameters. Average $\Tone$ measured throughout these runs and corresponding average $\FClTone$ are also listed.}
\end{table}

The robustness of the optimization protocol is tested by interleaving tuneups with CRB and $\Tone$ measurements over 11 hours (summarized in \cref{tab:verification}, and detailed in~\cite{SOMrestless}). Both tuneups reliably converge to $\FCl = 0.9991$, close to the $\Tone$ limit~\cite{MagesanNote}:
\begin{equation}
\label{eq:FTone}
\FClTone\approx \tfrac{1}{6}{\left(3 + 2 e^{-\tauc/2 \Tone} + e^{- \tauc/{\Tone}}\right)} = 0.9994,
\end{equation}
with $\tauc=1.875~\taup$. However, restless tuneup converges in one minute while  conventional tuneup requires eleven.

It remains to test how restless tuneup behaves as additional parameters are introduced.
Many realistic scenarios also require tuning the drive frequency $f$.
As a worst case, we take an initial detuning of $\pm250~\kHz$. The initial stepsize in the first (second) step is $100~\kHz$ ($50~\kHz$).
The 3-parameter optimization converges to $\FCl =0.9990\pm0.0001$ for both restless and conventional tuneups.
We attribute the slight decrease in $\FCl$ achieved by 3-parameter optimization to the observed reduction in average $\Tone$.

In summary, we have developed an accurate and robust tuneup method achieving a tenfold speedup over the state of the art~\cite{Kelly14}. This speedup is achieved by avoiding qubit initialization by relaxation and using real-time correlation of measurement outcomes to build the cost function for numerical optimization.
We have applied the restless concept to the tuneup of Clifford gates on a transmon qubit, reaching a $\Tone$-dominated fidelity of $0.999$ in one minute, verified by conventional randomized benchmarking and gate set tomography.
We have shown experimentally that the method can detect fractional reductions in gate error with nearly constant signal-to-noise ratio. Immediate next experiments will extend the restless concept to the tuneup of two-qubit gates and measurement operations, and to simultaneous tuneup of the physical qubits comprising a logical qubit.

\begin{acknowledgments}
We thank R.~Sagastizabal for experimental assistance, C.~Dickel, J.~Helsen, and S.~Poletto for discussions, A.~Johnson for support with Microsoft QCoDeS and K.~Ruddinger, E.~Nielsen and R.~Blume-Kohout for support with GST/pyGSTi.
This research is supported by the Office of the Director of National Intelligence (ODNI), Intelligence Advanced Research Projects Activity (IARPA),
via the U.S. Army Research Office grant W911NF-16-1-0071. Additional funding provided by the ERC Synergy Grant QC-lab, the China Scholarship Council (X.F.) and Microsoft Corporation Station~Q.
The views and conclusions contained herein are those of the authors and should not be interpreted as necessarily representing the official policies or endorsements, either expressed or implied, of the ODNI, IARPA, or the U.S. Government. The U.S. Government is authorized to reproduce and distribute reprints for Governmental purposes notwithstanding any copyright annotation thereon.
\end{acknowledgments}

\bibliographystyle{apsrev4-1}

\clearpage
\onecolumngrid
\renewcommand{\theequation}{S\arabic{equation}}
\renewcommand{\thefigure}{S\arabic{figure}}
\renewcommand{\thetable}{S\arabic{table}}
\renewcommand{\bibnumfmt}[1]{[S#1]}
\renewcommand{\citenumfont}[1]{S#1}
\setcounter{figure}{0}
\setcounter{equation}{0}
\setcounter{table}{0}
\section*{Supplemental material for ``Restless Tuneup of High-Fidelity Qubit Gates''}
\date{\today}
\maketitle
This supplement presents the hardware configuration used for the numerical tuneup, the characterization and modeling of the signal and noise of restless randomized benchmarking, and the procedure for calculating Clifford gate fidelities from GST process matrices. Finally, it presents the data summarized in Table~1 of the main text.

\section{Setup for numerical optimization}

\begin{figure*}[h]
  \centering
    \includegraphics[width=0.5\textwidth]{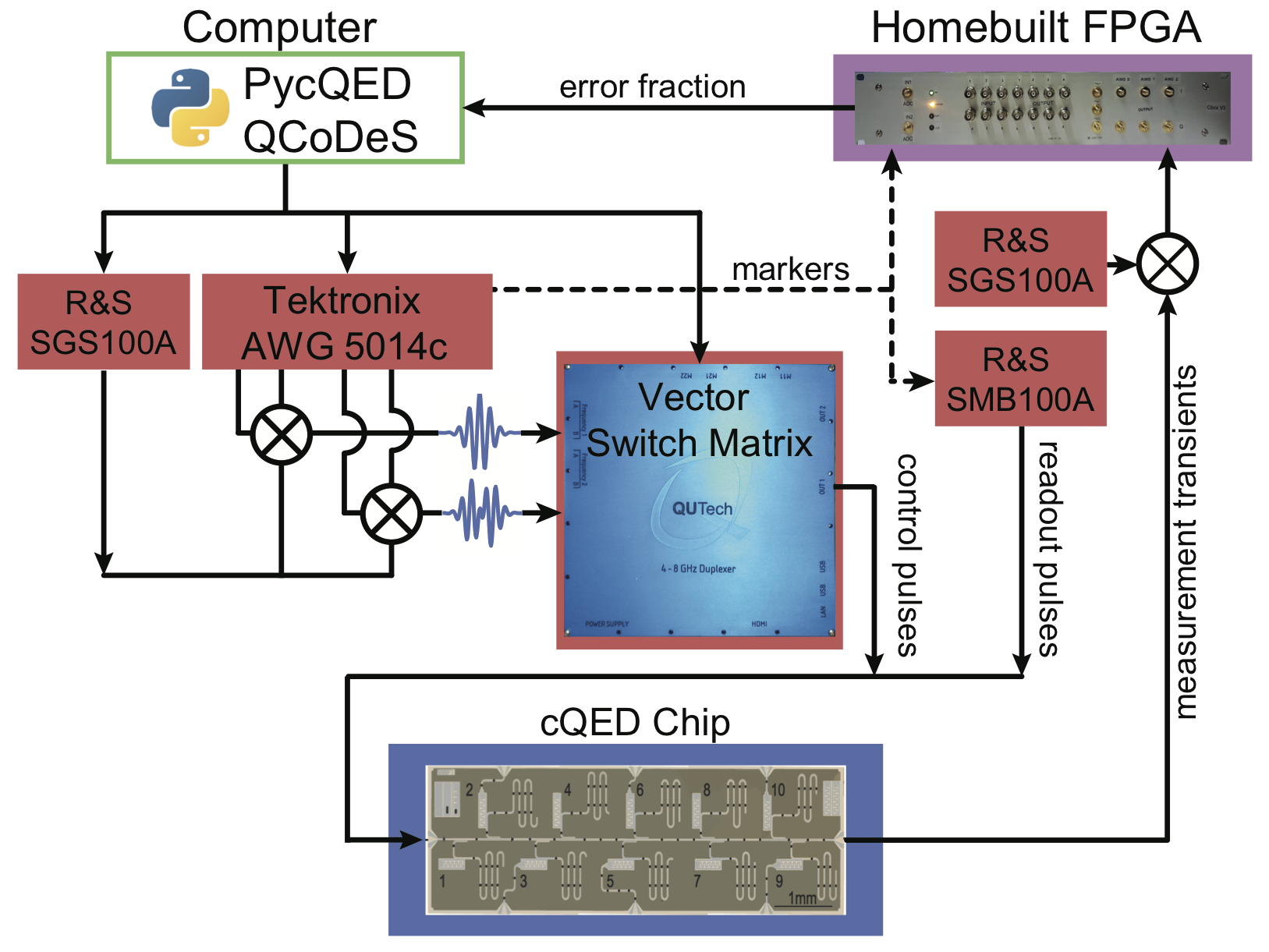}
    \caption{
    Schematic overview of the hardware components used in the numerical tuneup.
    }
    \label{fig:closed_loop}
\end{figure*}

The key hardware components executing the tuneup loop of Fig.~1(a)  are shown in \cref{fig:closed_loop}.
The computer is responsible for preparing the experiment and executing the numerical algorithm determining the parameter values for each iteration.
To do this, the computer relies on two python packages, \emph{PycQED} for cQED-specific routines~\cite{PycQED16} and \emph{QCoDeS} for the framework of instrument drivers~\cite{QCoDeS16}.
Part of the preparation consists of generating and uploading a sequence of control pulses and markers to the AWG.
Once an experiment starts, the AWG is responsible for all time-critical matters, including gating the readout pulses on the microwave source and triggering the data acquisition on the FPGA controller.
The control pulses are generated using 4 AWG channels, 2 for the $I$ and $Q$ quadratures of the Gaussian component and 2 for the quadratures of the derivative component.
The components are upconverted using single-sideband mixers and a constant microwave tone as a local oscillator (LO).
This allows independent control over the amplitude of both pulse components, using either the AWG or the VSM.
The frequency of the pulses can be changed by changing the frequency of the LO.
Note that all these controls can be applied without regenerating and uploading the sequence of control pulses to the AWG.
The transmon is read-out by interrogating its dispersively coupled resonator near its fundamental frequency using a capacitively coupled feedline.
Readout transients are amplified at the front end of the amplification chain by a Josephson parametric amplifier operated in the non-degenerate mode, providing $14~\dB$ of gain.
The FPGA controller performs final demodulation, integration and discrimination of measurement transients and real-time calculation of $\errfrac$.

\section{Signal and noise of the restless cost function}
\label{sec:SNR}
We experimentally obtained the signal and noise of RRB presented in Fig.~3 of the main text from 50 RRB experiments ($N=8000$ measurement outcomes each) at each $\NCl$ (32 values) and $\FCl$ (5 values). Here, $\FCl$ was varied by changing $\AG$.
The procedure was repeated 10 times for all settings to build up statistics. In this section, we present the derivation of the extended model used to predict these curves~(\cref{sec:theory_fluctuations}), using independent measurements of qubit $\Tone$ fluctuations performed one day apart~(\cref{sec:T1_fluctuations_experiment}).

\subsection{Modelling}
\label{sec:theory_fluctuations}

We develop a model for the RRB experiment to capture both the signal and noise obtained experimentally.
The standard deviation differs from that simply expected from a binomial distribution.
This is hypothesized to be caused by $\Tone$ fluctuations that are quasi-static during individual RRB experiments but dynamic on the time scale required for 50 repetitions.
We attempt to match the experimental results with a model containing $\Tone$ and its fluctuations, a relaxation independent pulse error $\ppulse$, and a SPAM offset $\psc$.
Independent measurements of the average and standard deviation of $\Tone$, and extractions of $\ppulse$ and $\psc$ from the data in Fig.~2(a) are used to produce the model curves in Fig.~3.

\subsubsection{Modeling without $\Tone$ fluctuations}
The time taken for a single-shot RRB experiment can be written $\trrb = \tro+ \tcl  \NCl$.
The static time $\tro=4.25~\us$ is the readout-and-depletion time, whilst the Clifford-dependent time $\tcl =37.5~\ns$ is the average time it takes to perform a Clifford gate.
To each of these we can associate an error rate, making the total error rate per single-shot experiment
\begin{equation*}
\pe=\ps~{+_p}~{\NCl}~{\times_p}~{\pcsupp}.
\end{equation*}
Here, $\ps$ is the error contribution due to SPAM, and $\pcsupp=1-\FCl$ is the error contribution per Clifford.
We must be careful with adding probabilities here, as two errors cancel.
This is taken care of by the probabilistic addition $a~{+_p}~b=a+b-2ab=a(1-b)+b(1-a)$, and the probabilistic multiplication $c~{\times_p}~a=a~{+_p}~a~{+_p}\ldots{+_p}a$ (repeated $c$ times for $c$ a positive integer). This multiplication can be simplified:
\begin{align*}
{\NCl}~{\times_p}~{\pcsupp} &=\NCl {\pcsupp} (1-{\pcsupp} )^{\NCl -1}+{\NCl \choose 3}{\pcsupp} ^3(1-{\pcsupp} )^{\NCl -3}+\ldots\\
&=\frac{1}{2}\left[((1-{\pcsupp} )+{\pcsupp} )^{\NCl} -((1-{\pcsupp} )-{\pcsupp} )^{\NCl} \right]\\
&=\frac{1}{2}\left[1-(1-2{\pcsupp} )^{\NCl} \right],
\end{align*}
resulting in a final error rate
\begin{align}
\label{eq:error_rate}
\pe=\ps+\frac{1}{2}[1-(1-2{\pcsupp} )^{\NCl} ](1-2\ps).
\end{align}

\subsubsection{Modelling with $\Tone$ fluctuations}
If $\ps$ or $\pcsupp$ fluctuate, the error rate $\pe$ for any given single-shot experiment is drawn from a distribution with mean
\begin{align*}
\overline{\pe}=\overline{\ps}+\frac{1}{2}[1-(1-2{\overline{\pcsupp}} )^{\NCl} ](1-2\overline{\ps}),
\end{align*}
and variance
\begin{align*}
\var(\pe)&=(1-2\overline{\pcsupp} )^{2\NCl}\var(\ps)+{\NCl}^2(1-2\overline{\ps})^2(1-2\overline{\pcsupp} )^{2(\NCl-1)}\var(\pcsupp)\\
&+2{\NCl}(1-2\overline{\ps})(1-2{\pcsupp} )^{2{\NCl}-1}\covar({\pcsupp} ,\ps).
\end{align*}
Here, $\covar(\pcsupp ,\ps)$ is the covariance between $\pcsupp$ and $\ps$, and $\overline{\pcsupp}$ [$\var(\pcsupp)$] and $\overline{\ps}$ [$\var(\ps)$] are the means [variances] of $\pcsupp$ and $\ps$, respectively.

Measurements of $\er$ use $N=8000$ single-shot measurement outcomes, which we assume are selected from a binomial distribution with mean $(1-P)$. $P$ is in turn selected from a distribution with mean $\overline{\pe}$ and standard deviation $\sigma_{\pe}$.
Let $N_e$ be the number of erroneous measurements, given as $N_e=N\er$.
In order to calculate the mean and variance in $N_e$, we have to calculate the first and second moments of the distribution, averaged over all $P$.
We assume a normal distribution for $P$. For the first moment we obtain
\begin{align*}
\langle N_e\rangle&=\int_{-\infty}^{\infty}\left[\sum_{k=0}^{N}k{N\choose k}P^k(1-P)^{N-k}\right]e^{-\frac{(P-\overline{\pe}))^2}{(2\sigma_{\pe}^2)}}\frac{1}{\sqrt{2\pi\sigma_{\pe}^2}}dP\\
&=N\int_{-\infty}^{\infty}Pe^{-\frac{(P-\overline{\pe})^2}{(2\sigma_{\pe}^2)}}\frac{1}{\sqrt{2\pi\sigma_{\pe}^2}}dP=N\overline{\pe}.
\end{align*}
As expected, the average number of erroneous measurements equals the total number of measurements multiplied by the average error, and is unaffected by fluctuations.
For the second moment we calculate
\begin{align*}
\langle N_e^2 \rangle&=\int_{-\infty}^{\infty}\left[\sum_{k=0}^{N}k^2{N\choose k}P^k(1-P)^{N-k}\right]e^{-\frac{(P-\overline{\pe})^2}{(2\sigma_{\pe})^2)}}\frac{1}{\sqrt{2\pi\sigma_{\pe}^2}}dP\\
&=\int_{-\infty}^{\infty}(NP+N(N-1)P^2)e^{-\frac{(P-\overline{\pe})^2}{(2\sigma_{\pe}^2)}}\frac{1}{\sqrt{2\pi\sigma_{\pe}^2}}dP\\
&=N\overline{\pe}+N(N-1)({\overline{\pe}}^2+\sigma_{\pe}^2).
\end{align*}
This leads to the final result:
\begin{equation}
\label{eq:noise_model}
\var(\er)=\frac{1}{N}\overline{\pe}(1-\overline{\pe})+\frac{N-1}{N}\var(\pe).
\end{equation}
The simple model without $\Tone$ fluctuations can be recovered here by setting $\var(\pe)=0$.

\subsubsection{Asymmetry}
Due to the asymmetry of $\Tone$, the error rate $\pej$ depends on whether the qubit is in the excited or ground state during $\tro$.
The measurement, lasting $\tm = 1~\us$, is $\Tone$ rather than noise limited. We can approximate it by perfect state update and measurement at $\tb \approx
4\tm/7 = 0.57~\us$~\cite{oBrien16}, followed by a rest time $\ta = \tro -\tb = 3.68~\us$ before the beginning of the next Clifford sequence.
Let the system state at the point of the measurement (i.e., $\tb$ into the measurement time) be $\ket{j}$ with $j=0$ or $1$.
If a single error occurs during the sequence, the flipping sequence will revert the qubit to the same state $\ket{j}$ at the next measurement point.
This implies that the process is biased towards states with higher error rate, and so the error rate cannot be simply averaged over that expected individually for $\ket{0}$ and $\ket{1}$.
Instead, we let the population fraction of $\ket{j}$ over the experiment be $f_j$,
and solve the steady-state rate equation for $f_j$:
\begin{equation*}
f_j=\pej f_j+(1-\pe^{(1-j)})(1-f_j).
\end{equation*}
This leads to an error rate of
\begin{equation}
\pe=\frac{\pe^{(0)}(1-\pe^{(1)})+\pe^{(1)}(1-\pe^{(0)})}{(1-\pe^{(0)})+(1-\pe^{(1)})}.
\label{eq:error_rate_asym}
\end{equation}
The error during the RRB sequence is state independent, and so the adjustment to \cref{eq:error_rate} comes solely from the adjustment to the SPAM error:
\begin{align*}
\pej=\psj+\frac{1}{2}[1-(1-2{\pcsupp} )^{\NCl} ](1-2\psj),
\end{align*}
with
\begin{align*}
\pszero=\psc+(1-e^{-\tb/\Tone}),\;\;\;\;
\psone=\psc+(1-e^{-\ta/\Tone})e^{-\tb/\Tone}.
\end{align*}
Here, $\psc$ is a small error accounting for non-$\Tone$ SPAM.
Substituting these into \cref{eq:error_rate_asym} allows for the calculation of the error $\pe$ as a function of $\pcsupp$, $\NCl$, and $\Tone$. In order to calculate the standard deviation, we must then calculate the first derivative, via
\begin{equation}
\label{eq:pe_derivative}
\frac{\partial\pe}{\partial\Tone}=\sum_j\frac{\partial\pe}{\partial\pej}\left(\frac{\partial\pej}{\partial\psj}\frac{\partial\psj}{\partial\Tone}+\frac{\partial\pej}{\partial\pcsupp}\frac{\partial\pcsupp}{\partial\Tone}\right).
\end{equation}
Here, the value of $\frac{\partial\pcsupp}{\partial\Tone}$ is obtained by assuming that $\pcsupp$ can be split into a constant pulse error probability $\ppulse$ plus a $\Tone$-induced error probability $\pctone=1-\FClTone$, with $\FClTone$ as defined in Eq.~(3).

\subsection{Measurement of $\Tone$ fluctuations}
\label{sec:T1_fluctuations_experiment}
\begin{figure}[htb]
  \centering
  \includegraphics{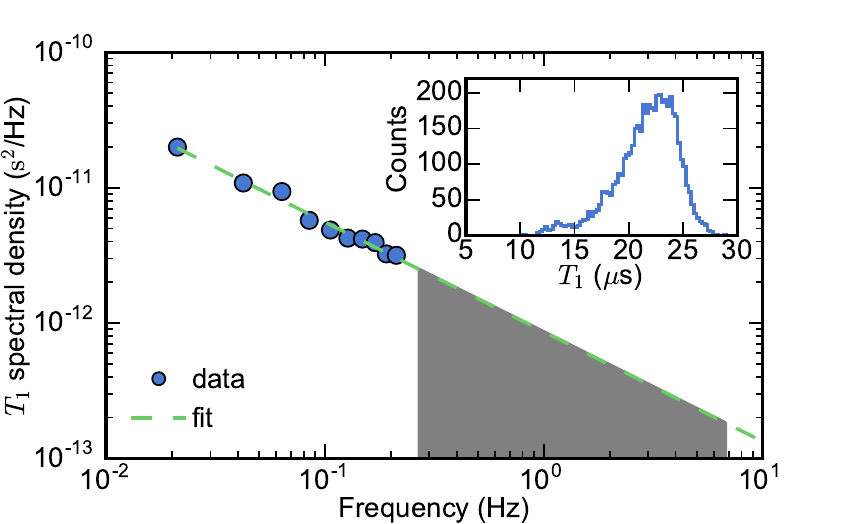}
  \caption{\label{fig:PSD}
    Power spectral density of $\Tone$ fluctuations.
    Main panel: measured single-sided PSD of $\Tone$ fluctuations and best fit (see details in text). The indicated frequency range is that relevant for estimating $\sigma_{\Tone}$ in the RRB experiments of \cref{sec:SNR}. Inset: Histogram of 4914 $\Tone$ measurements. The set has $\overline{\Tone} = 21.6~\us$.
    }
\end{figure}

We perform repeated measurements of $\Tone$ one day after the RRB experiments.
We extract $\Tone$ from exponential best fits to standard sliding $\pi$-pulse experiments. These measurements rely on qubit initialization by waiting.
The benefit of this method is that one can measure $\Tone$ fluctuations independently from fluctuations in residual qubit populations, gate fidelity and readout fidelity (unlike restless sequences).
The downside is that one can only probe $\Tone$ in $\Delta t =2.0~\s$ intervals.
We measure $\Tone$ in  $L=234$ runs $l$ of $M=21$ measurements each, and calculate the single-sided power spectral density (PSD) as
\[
\SD (f) = \frac{2\Delta t}{LM} \sum_{l=1}^L\left|\sum_{m=1}^M \delta T_{1,l} [m] e^{-i2\pi f m \Delta t }\right|^2,
\]
where $\delta T_{1,l}[m]= T_{1,l}[m]-\frac{1}{M} \sum_{m'=1}^{M} T_{1,l}[m']$.
We fit  $\SD (f)= \alpha \left( f / 1~\Hz\right)^\beta$ to the experimental PSD, finding best-fit parameters $\alpha=8.4\cdot10^{-13}~\s^2/\Hz$ and $\beta=-0.81$ (data and fit are shown in \cref{fig:PSD}).
Extrapolating the PSD to higher frequencies, we can estimate the expected $\sigma_{\Tone}$ in the RRB experiments of \cref{sec:SNR} by integrating over the frequency interval bounded above by the rate of single RRB experiments ($f_\mathrm{u}=1/0.074~\s$ at low $\NCl$) and below by the acquisition time for 50 such experiments ($f_\mathrm{l}=1/3.7~\s$).
We find $\overline{\Tone}=21.6~\us$ and
\[
\sigma_{\Tone} = \left( \int_{f_\mathrm{l}}^{f_\mathrm{u}} \SD df  \right)^{1/2} = 2.44\pm0.1~\us.
\]
We estimate the uncertainty in $\sigma_{\Tone}$ by splitting the dataset into 6 subsets of equal length.

\subsection{Relation to experiment}
Using the measured $\overline{\Tone}$, we fit \cref{eq:error_rate_asym} to the data in Fig.~2(a) to extract a common $\psc=0.006$ and curve specific $\ppulse$.
We use \cref{eq:error_rate_asym,eq:pe_derivative} to obtain the model curves for $\Delta\overline{\er}$ and $\overline{\sigma_{\er}}$ shown in Fig. 3 of the main text, finding good agreement with experiment.

\section{Gate Set Tomography and Randomized Benchmarking Fidelities}
In order to compare results from GST to those acquired using CRB, the results of GST need to be converted to Clifford fidelities.
GST performs a full self-consistent tomography of the gates in the set  $\{I, X90, Y90, X180, Y180\}$, consisting of the identity and positive $\pi/2$ and $\pi$ rotations around the $x$ and $y$ axes.
The super-operators for the gates in the gate set are extracted from the GST data using pyGSTi~\cite{Nielsen16}.
These are then used to construct the 24 elements ($G_{\mathrm{Cl}_n}^{\mathrm{GST}}$) of the (single-qubit) Clifford group ($\Clgroup$) according to the decomposition of~\cite{Epstein14_SOM}.
To account for the missing negative rotations in the gate set, we replace negative rotations with their positive counterparts (e.g., $-X90 \rightarrow X90$)
For each of these operations, the depolarization probability is calculated by looking at the overlap with the target state ($|\rho_t\rangle \rangle$ in the super-operator formalism)
after applying $G_{\mathrm{Cl}}^{\mathrm{GST}}$ to the input state $|{\rho_i}\rangle \rangle$, for all poles of the Bloch sphere as input states and taking the geometric mean:
\[
p_n = \sqrt[6]{\prod_{\rho_i}{\langle\bra{\rho_\mathrm{t}} G_{\mathrm{Cl-}n}^{\mathrm{GST}} \ket{\rho_\mathrm{i}}\rangle}},
\]
where the target state is the state one would get if the gates were perfect:
\[
\ket{\rho_\mathrm{t}}\rangle=G_{\mathrm{Cl-}n}^{\mathrm{Ideal}} \ket{\rho_\mathrm{i}}\rangle.
\]
$\pc$ is the geometric mean of the individual depolarization probabilities for all $G_\mathrm{Cl_n} \in \Clgroup$ and related to $\FCl$ through $\FCl  =\frac{1}{2}+\frac{1}{2} \pc$.

\cref{tab:GST} summarizes the gate fidelities found after performing the two-parameter optimization, for the four starting ($\AG$,~$\AD$) conditions discussed in the main text.

\begin{table}
    \begin{tabular}{ c| c c }
         & Conventional  & Restless  \\ \hline \hline
        $F_{I}$       &  $0.99928\pm0.00007$ &    $0.99921\pm0.00005$\\
        $F_{X90}$     &  $0.99927\pm0.00005$ &    $0.99925\pm0.00004$\\
        $F_{X180}$    &  $0.99920\pm0.00007$ &    $0.99910\pm0.00005$\\
        $F_{Y90}$     &  $0.99908\pm0.00005$ &    $0.99906\pm0.00005$\\
        $F_{Y180}$    &  $0.99901\pm0.00008$ &    $0.99891\pm0.00005$\\
        \hline
         $\FClGST$  & ${0.99909\pm 0.00005}$ &    ${0.99907\pm 0.00003}$    \\
         $\FCl$ & ${0.9991}$ & {${0.9991}$}                              \\ \hline
    \end{tabular}
    \caption{
        \label{tab:GST}
        Measured gate fidelities in GST.
        Gate fidelities correspond to average gate fidelities for the four starting conditions of the two-parameter optimization as discussed in the main text.}
\end{table}

\section{Verification of conventional and restless tuneup}
\begin{figure}[!htb]
  \centering
  \includegraphics{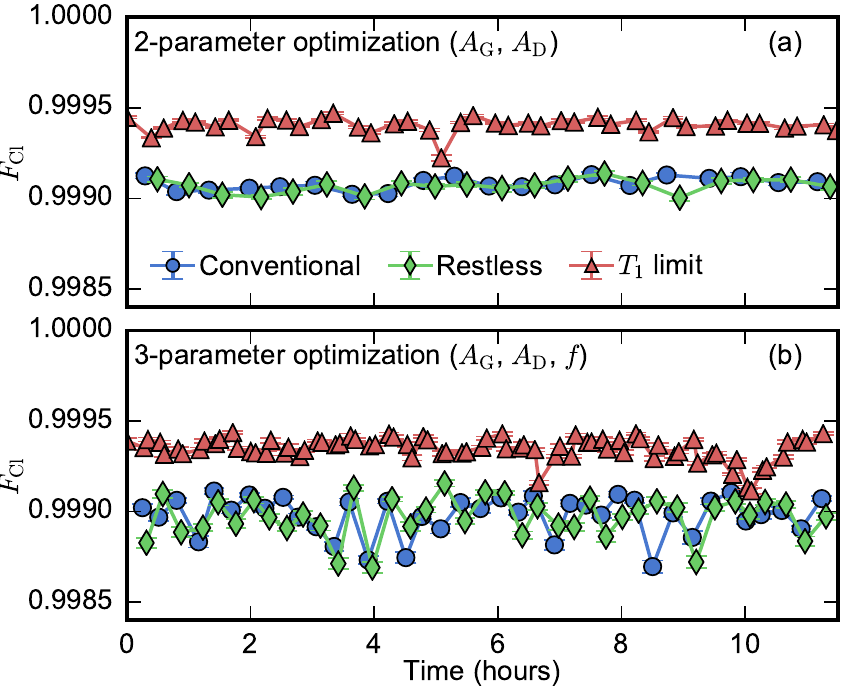}
  \caption{
  Performance comparison of repeated restless and conventional tuneups for two parameters (a) and three parameters (b).
  Each iteration consists of a conventional tuneup followed by a CRB measurement of $\FCl$, a restless tuneup followed by a CRB measurement of $\FCl$, and a $\Tone$ experiment to determine $\FClTone$.
  For each iteration, a new starting condition is chosen (detailed in main text) that is used for both the conventional and restless tuneup.
}
\label{fig:verification}
\end{figure}

The speed, robustness and accuracy of the two- and three- parameter optimizations are tested during an 11-hour period by interleaving conventional and restless tuneups with CRB and $\Tone$ experiments.
The data summarized in Table~1 of the main text is shown in~\cref{fig:verification}.
The two-parameter (three-parameter) optimization loops over 4~(8) different starting conditions as specified in the main text.
The starting condition is updated after each set of conventional and restless optimizations.

\bibliographystyle{apsrev4-1}

\end{document}